\documentclass[sigplan,review,anonymous]{acmart}

\usepackage{booktabs}
\usepackage{adjustbox}
\usepackage{amsmath}
\usepackage{amsfonts}
\usepackage{listings}
\usepackage{color}
\usepackage{xspace}
\usepackage{comment}
\usepackage{graphicx} 
\usepackage{appendix}
\usepackage{wrapfig}
\usepackage{multirow}
\usepackage{bm}
\usepackage{microtype}
\usepackage{enumitem}
\usepackage{marginalia}

\setlist{leftmargin=5.5mm}

\definecolor{listinggreen}{rgb}{0,0.6,0}
\definecolor{listinggray}{rgb}{0.5,0.5,0.5}
\definecolor{listingmauve}{rgb}{0.58,0,0.82}
\definecolor{listingkeywordcolor}{rgb}{1.0,0.4,0.0}
\definecolor{listinglightgray}{rgb}{0.8863,0.8863,0.8863}

\newcommand\yes{{\tiny Yes}}
\newcommand\no{\color{red}{{\tiny No}}}

\newcommand\COMMENT[1]{}
\newcommand\framework{\textsc{Tiramisu}\xspace}

\newcommand{\HIDE}[1]{}
\newcommand\TODO[1]{\mbox{\textcolor{red}{#1}}}

\newif\iflong

\acmConference[PL'20]{ACM SIGPLAN Conference on Programming Languages}{June 15--20, 2020}{London, United Kingdom}
\acmYear{2020}
\acmISBN{} 
\acmDOI{} 
\startPage{1}

\setcopyright{none}

\usepackage{booktabs}   
\usepackage{subcaption} 

\begin{document}

\title[\framework{} Compiler for DNNs]{\framework{}: A Polyhedral Compiler for Dense and Sparse Deep Learning}         


\author{First1 Last1}
\authornote{with author1 note}          
\orcid{nnnn-nnnn-nnnn-nnnn}             
\affiliation{
  \position{Position1}
  \department{Department1}              
  \institution{Institution1}            
  \streetaddress{Street1 Address1}
  \city{City1}
  \state{State1}
  \postcode{Post-Code1}
  \country{Country1}                    
}
\email{first1.last1@inst1.edu}          

\author{First2 Last2}
\authornote{with author2 note}          
\orcid{nnnn-nnnn-nnnn-nnnn}             
\affiliation{
  \position{Position2a}
  \department{Department2a}             
  \institution{Institution2a}           
  \streetaddress{Street2a Address2a}
  \city{City2a}
  \state{State2a}
  \postcode{Post-Code2a}
  \country{Country2a}                   
}
\email{first2.last2@inst2a.com}         
\affiliation{
  \position{Position2b}
  \department{Department2b}             
  \institution{Institution2b}           
  \streetaddress{Street3b Address2b}
  \city{City2b}
  \state{State2b}
  \postcode{Post-Code2b}
  \country{Country2b}                   
}
\email{first2.last2@inst2b.org}         

\begin{abstract}


In this paper, we demonstrate a compiler that can optimize {\em sparse and recurrent neural networks}, both of which are currently outside of the scope of existing neural network compilers (sparse neural networks here stand for networks that can be accelerated with sparse tensor algebra techniques).
Our demonstration includes a mapping of sparse and recurrent neural networks to the polyhedral model along with an implementation of our approach in \framework{}, our state-of-the-art polyhedral compiler. We evaluate our approach on a set of deep learning benchmarks and compare our results with hand-optimized industrial libraries.  Our results show that our approach at least matches Intel MKL-DNN  and in some cases outperforms it by $5\times$ (on multicore-CPUs).
\end{abstract}



\maketitle

\vspace{-0.5cm}
\section{Introduction}
\label{sec:intro}
\vspace{-0.25cm}

With the increasing need for efficient deep learning, there is a surge in hardware and compiler research, not only because compilers improve developer productivity by generating code for the new deep learning hardware accelerators, but also because compilers can significantly optimize deep learning computations (e.g., through operator fusion~\cite{Vasilache2018TensorCF}).

Generating high performance code for deep learning requires complex code and data layout transformations, management of complex memory hierarchies, and the ability to take advantage of complex low level hardware features.
While state-of-the-art deep learning compilers can optimize efficiently neural networks with acyclic data-flow graphs (feed-forward neural networks), they still have limitations in optimizing recurrent and sparse neural networks.

In this paper, we demonstrate a compiler that can optimize {\em sparse and recurrent neural networks}~\footnote{Sparse neural networks in this context mean neural networks that can be accelerated with sparse tensor algebra techniques}.
We implement our approach in \framework{}~\cite{Baghdadi:2019:TPC:3314872.3314896}, our state-of-the-art polyhedral compiler.
\framework{} takes a high level representation of the program (pure algorithm and a set of scheduling commands), applies the necessary code transformations, and generates highly-optimized code for the target architecture.
It uses the polyhedral representation internally, which provides many advantages such as the ability to apply complex loop and data layout transformations and the ability to express programs that have non-rectangular iteration spaces or that have cycles in their data flow graphs. \framework{} relies on the use of scheduling commands, therefore it avoids many limitations that fully automatic compilers have.
\framework{} has two unique features in the area of deep learning: (1) it introduces the first DNN (Deep Neural Network) compiler that exploits weight sparsity; and (2) it can express and optimize general RNNs (Recurrent Neural Networks).
In this paper, we will demonstrate \framework{} by generating code for multicore CPUs.

Exploiting weight sparsity in deep neural networks (DNNs) is a promising direction for accelerating deep learning. The weights of a neural network can be made sparse using network pruning~\cite{lecun_optimal_1990, han_learning_2015}, a technique to sparsify neural networks by removing unnecessary structure from the neural network while minimizing the loss in accuracy. Two families of network pruning techniques exist: pruning to obtain structured sparsity (e.g., by dropping convolutional filters~\cite{li_pruning_2017}) and pruning to obtain unstructured sparsity (e.g., by dropping individual weights or connections in the neural network~\cite{han_learning_2015}).
While structured sparsity is easy to accelerate, unstructured sparsity techniques can find much sparser networks with equivalent accuracy.
State-of-the-art unstructured network pruning techniques~\cite{he_amc_2018, gale_state_2019} can prune a ResNet-50 trained on ImageNet by $80\%$ without any loss in accuracy and a VGG-19 trained on CIFAR-10 by $99\%$~\cite{frankle_stabilizing_2019}.
State-of-the-art DNN compilers however do not exploit such unstructured sparsity, due to fine-grained sparsity patterns being more difficult to accelerate, and therefore do not realize the performance gains from reduced computation and memory accesses.

In this paper, we make the following \emph{contributions}:
\begin{itemize}
    \item We introduce the first DNN compiler that generates efficient code for neural networks with sparse weights; In particular, \framework{} is the first to show that deep neural networks with unstructured weight sparsity can be accelerated by compilers;
    \item We introduce 
    a DNN compiler that can express and optimize the general form of RNNs (where the number of RNN unrolling factor is unknown at compile time);
    \item We evaluate our compiler on a set of deep learning benchmarks and compare it with the Intel MKL-DNN library (on multicore-CPU).  We show that \framework{} can generate efficient code that matches or outperforms Intel MKL-DNN by up to $5\times$.
\end{itemize}

\vspace{-0.5cm}
\section{The \framework Embedded DSL}
\vspace{-0.25cm}

\framework{} is a domain-specific language (DSL) embedded in C++. It provides a pure C++ API that allows users to write a high level, architecture-independent algorithm and a set of scheduling commands that guide code generation.
\framework{} is integrated in high level deep learning frameworks such as Pytorch and therefore can be used transparently by end-users. It can also be generated by any other similar high level framework or DSL. 


The first part of a \framework{} program specifies the algorithm without specifying loop optimizations (when and where the computations occur) or data-layout (how data should be stored in memory).
The second part of the program provides the schedule, which specifies how the program should be optimized (vectorization, tiling, fusion, ...) and how the results of computations should be stored. The following code shows an example of a convolution algorithm written in \framework. 

\begin{lstlisting}[language=C,escapechar=@]
// Declare the iterators.
var n(0, batch), fout(0, out_features), fin(0, in_features);@\label{fig:example:tiramisu:iterators}@
var y(1, H-1), x(1, W-1), k0(0, 3), k1(0, 3);

// Algorithm.
conv(n, fout, y, x) +=
    weights(fout, fin, y, x) * input(n, fin, y+k0, x+k1);@\label{fig:example:tiramisu:computation1}@
\end{lstlisting}
\vspace{-0.1cm}

The iterators in line~\ref{fig:example:tiramisu:iterators} define the iteration domain of \texttt{conv} (i.e., loop bounds). The algorithm is semantically equivalent to the following code.

\begin{lstlisting}[language=C,escapechar=@]
for (n in 0..batch)
 for (fout in 0..out_features)
  for (y in 1..H-1)
   for (x in 1..W-1)
    for (fin in 0..in_features)
     for (k0 in 0..3)
      for (k1 in 0..3)
       conv[n, fout, y, x] += weigths[fout, fin, y, x] * input[n, fin, y+k0, x+k1];
\end{lstlisting}

The following code shows an example of scheduling commands (optimization commands) that can be applied on the previous convolution kernel. These commands parallelize the loop \texttt{n}, interchange the loops \texttt{fin} and \texttt{fout} and vectorize the loop \texttt{fout} by a vector length of 8.

\begin{lstlisting}[language=C,escapechar=@]
conv.parallelize(n);
conv.interchange(fin, fout);
conv.vectorize(fout, 8);
\end{lstlisting}

\vspace{-0.35cm}
\paragraph{Neural Network Optimizations}
Neural network optimizations applied by \framework{} include operator fusion, loop skewing, parallelization, multi-level tiling, loop reordering, loop unrolling, vectorization, array packing~\cite{Goto:2008:AHM:1356052.1356053}, register blocking, data prefetching, full/partial tile separation and tuning optimization parameters to the target architecture (e.g., choosing tile sizes or loop unrolling factors that are optimal for the target machine using auto-tuning~\cite{opentuner}).

\framework{} has two unique neural network optimizations: (1) optimizing sparse convolutions (weight sparsity); and (2) optimizing RNNs (Recurrent Neural Networks). In the next section we will provide more details about how does \framework{} support these two optimizations.

\vspace{-0.25cm}
\section{Optimizing Sparse Neural Networks}
\vspace{-0.25cm}

\begin{table*}[h!]
\scriptsize
\begin{adjustbox}{center}
    \centering
    \begin{tabular}{cccccccccccccccccccc}
    \toprule
     \textbf{Network} & \textbf{1} & \textbf{2} & \textbf{3} & \textbf{4} & \textbf{5} & \textbf{6} & \textbf{7} & \textbf{8} & \textbf{9} & \textbf{10} \\
     \midrule
     \textbf{VGG-16} & $49.5\%$ & $34.6\%$ & $77.7\%$ & $79.5\%$ & $77.1\%$ & $65.9\%$ & $45.7\%$ & $24.2\%$ & $5.8\%$ & $1.0\%$ \\
     \textbf{ResNet-20} & $61.3\%$ & $22.2\%$ & $24.0\%$ & $23.8\%$ & $21.3\%$ & $27.6\%$ & $19.4\%$ & $26.8\%$ & $20.3\%$ & $16.1\%$ \\
     \midrule
      \textbf{Network} & \textbf{11} & \textbf{12} & \textbf{13} & \textbf{14} & \textbf{15} & \textbf{16} & \textbf{17} & \textbf{18} & \textbf{19} \\
     \midrule
     \textbf{VGG-16} & $0.2\%$ & $0.2\%$ & $0.3\%$ & $0.4\%$ & $0.7\%$ & $1.0\%$ & N/A & N/A & N/A \\
     \textbf{ResNet-20} & $12.4\%$ & $16.3\%$ & $11.0\%$ & $15.7\%$ & $13.0\%$ & $11.3\%$ & $9.2\%$ & $10.0\%$ & $2.1\%$ \\
     \bottomrule
    \end{tabular}
    \end{adjustbox}
    \caption{Density across conv layers in a pruned ResNet-20 and VGG-16}
    \label{tab:layerwise-densities}
    \vspace{-0.25cm}
\end{table*}

Modern CNNs for vision tend to be significantly overparameterized, imposing much higher memory and computational requirements than necessary for the task~\citep{han_learning_2015}.
However, it is typically not possible to simply reduce the model size by using smaller models to begin with: small models trained from scratch do not reach the same accuracy as large models which are trained then sparsified~\citep{zhu_prune_2018}.
Instead, the smallest models are obtained through \emph{unstructured} pruning techniques: training a full model, then pruning individual weights from that model using some heuristic in order to create the most accurate model at a given sparsity level~\citep{han_learning_2015}.

In this paper, we evaluate on networks obtained through a technique based on the Lottery Ticket Hypothesis~\citep{frankle_stabilizing_2019} (although support for sparse weights in \framework{} is general and does not depend on the patterns of sparsity produced by the Lottery Ticket Hypothesis work).
This technique iteratively trains a network, prunes it by simply removing the $20\%$ of weights with the lowest magnitude throughout the network, rewinds the weights to their values early in training, then re-trains and repeats.
Using this technique results in sparse networks that reach the same accuracy as the original dense network: we can prune a ResNet-20 to $21\%$ density and a VGG-16 to only $1\%$ density without any loss in accuracy.
However, these sparse networks are not uniformly sparse across all layers: early layers (with few channels, and therefore few parameters) tend to be minimally pruned and end up dense. However later layers (with many channels and are correspondingly larger) tend to be pruned to be sparser.
The layerwise sparsity rates for ResNet-20 and VGG-16 are presented in Table~\ref{tab:layerwise-densities}.

\vspace{-0.25cm}
\paragraph{Sparse Convolution with CSR}
The following code shows the algorithm that we use to implement convolutions that exploit weight sparsity~\cite{park2016faster}. We store the weight tensors in a CSR-like format (Compressed Sparse Row). This format is created as follows: first, we flatten the original weight tensor which has the following dimensions (OutputFeatures, InputFeatures, K, K)~\footnote{k is the size of the convolution filter (e.g., $3\times 3$)} to (OutputFeatures, InputFeatures$\times$K$\times$K); then we compress the rows of the resulting matrix using CSR.

\begin{lstlisting}[language=C,escapechar=@]
for each output channel n
  for j in (W.rowptr[n], W.rowptr[n+1]) {
    off = W.colidx[j]; coeff = W.value[j];
    for (int y = 0; y < H_OUT; ++y)
      for (int x = 0; x < W_OUT; ++x)
        out[n][y][x] += coeff*in[y*W_OUT+x+off)]
  }
\end{lstlisting}


\vspace{-0.5cm}
\section{Expressing and Optimizing Recurrent Neural Networks}
\vspace{-0.25cm}
Many state-of-the-art DNN compilers do not allow users to
express dynamic RNNs. Halide~\cite{halide_12}, for example,
is designed to express programs with acyclic dependence
graphs (which excludes dynamic RNNs);
this restriction is imposed by the Halide language and compiler 
to guarantee the correctness of optimizations.
To avoid this overconservative language restriction, \framework{}
relies on dependence analysis instead to check for the correctness of code
transformations, enabling the user to express dynamic RNNs and optimize them.

In order to parallelize the execution of multilayer-LSTMs, \framework{}
applies a transformation known as iteration space skewing which exposes
wavefront parallelism hidden in multilayer-LSTMs. Such parallelization
is necessary for increasing GPU occupancy when targeting GPUs, it is also
necessary to parallelize multilayer-LSTMs when targeting distributed
architectures.

\section{Support for RNNs}
\TODO{Show example (graphic) that cuDNN and Halide cannot do}
\\
\TODO{Explain why cuDNN and Halide cannot support it}
\\
\TODO{Show how we can optimize it (graphic)}

\section{Compiler Implementation}

\subsection{Support for Sparse DNNs}
A key challenge with representing Sparse DNNs is that sparse implementations require non-affine loop bounds and array accesses, which do not fit within the polyhedral model's traditional affine requirements. To support Sparse DNNs, we therefore extend the polyhedral model to support non-affine loop bounds and array accesses in a way similar to ~\cite{benabderrahmane_polyhedral_2010}.

\TODO{Specify loop bounds to support sparsity}

\subsection{Support for RNNs}
While frameworks such as Halide~\cite{halide_12} and Tensor Comprehensions~\cite{DBLP:journals/corr/abs-1802-04730} can implement LSTMs (Long Short Term Memory), they are limited to LSTMs where the number of \emph{recurrence steps} (i.e., cells) is known at compile time. This is mainly because Halide and Tensor Comprehensions have a restricted language that prevents the user from writing a program that has a cycle in its data flow graph since they cannot guarantee the correctness of loop optimizations in that case.
Unlike Halide and Tensor Comprehensions, \framework{} performs polyhedral dependence analysis~\cite{feautrier_dataflow_1991} and uses transformation legality checks to guarantee the correctness of code transformations. This gives the compiler more flexibility and removes an important restriction on the \framework{} language.

Parallelizing LSTMs is a diffcult task. LSTM cells can be stacked on top of each other, where the output of one cell becomes the input of the next. Due to the sequential data flow in LSTM, the cells of a single layer network cannot run in parallel. But, in fact, there is hidden parallelism that can be extracted using \framework{} iteration space skewing is applied (to extract wavefront parallelism). 

\vspace{-0.25cm}
\section{Evaluation}
\vspace{-0.25cm}

We evaluate \framework{} on a set of deep learning benchmarks. We compare it with the Intel MKL-DNN (1.0) and cuDNN (7.0) libraries which provide highly optimized implementations for Intel multicore CPUs and Nvidia GPUs.

The CPU evaluation is performed on an 8-core Intel i7-6700HQ CPU, 16 GB RAM, Ubuntu 18.04. The GPU evaluation is performed on an Nvidia Pascal P4 GPU. Each experiment is repeated $30\times$ and the median time is reported.

\begin{figure}[h!]
\vspace{-0.5cm}
\centering
    \hspace{-2cm}
    \begin{minipage}{0.6\textwidth}
        \includegraphics[scale=0.28]{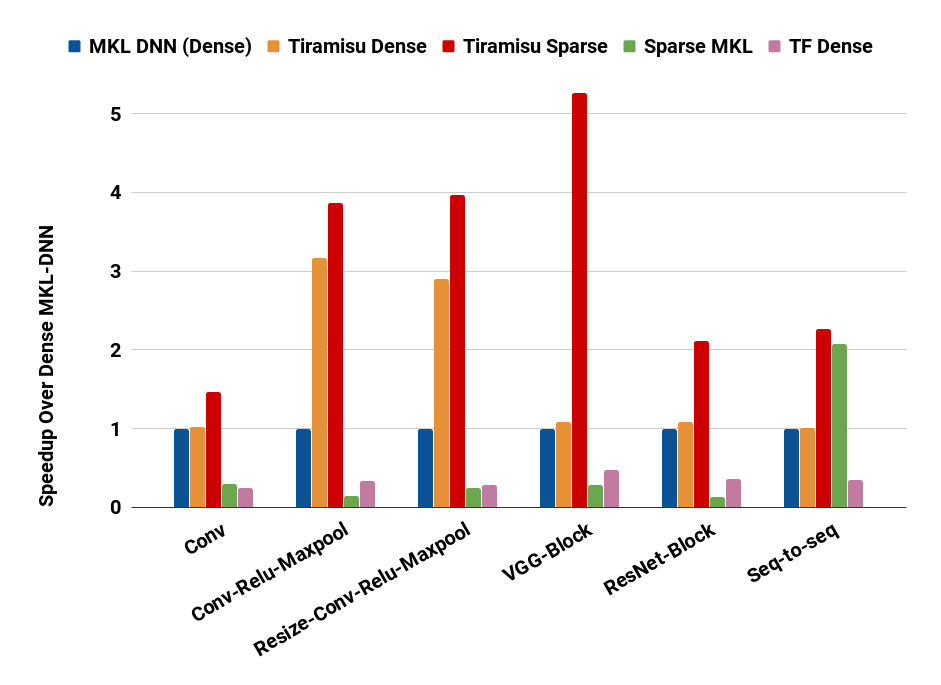}
        \caption{Speedups over Intel MKL-DNN.}
        \label{DNNBenchmark}
    \end{minipage}
    \begin{minipage}{0.3\textwidth}
        \includegraphics[scale=0.25]{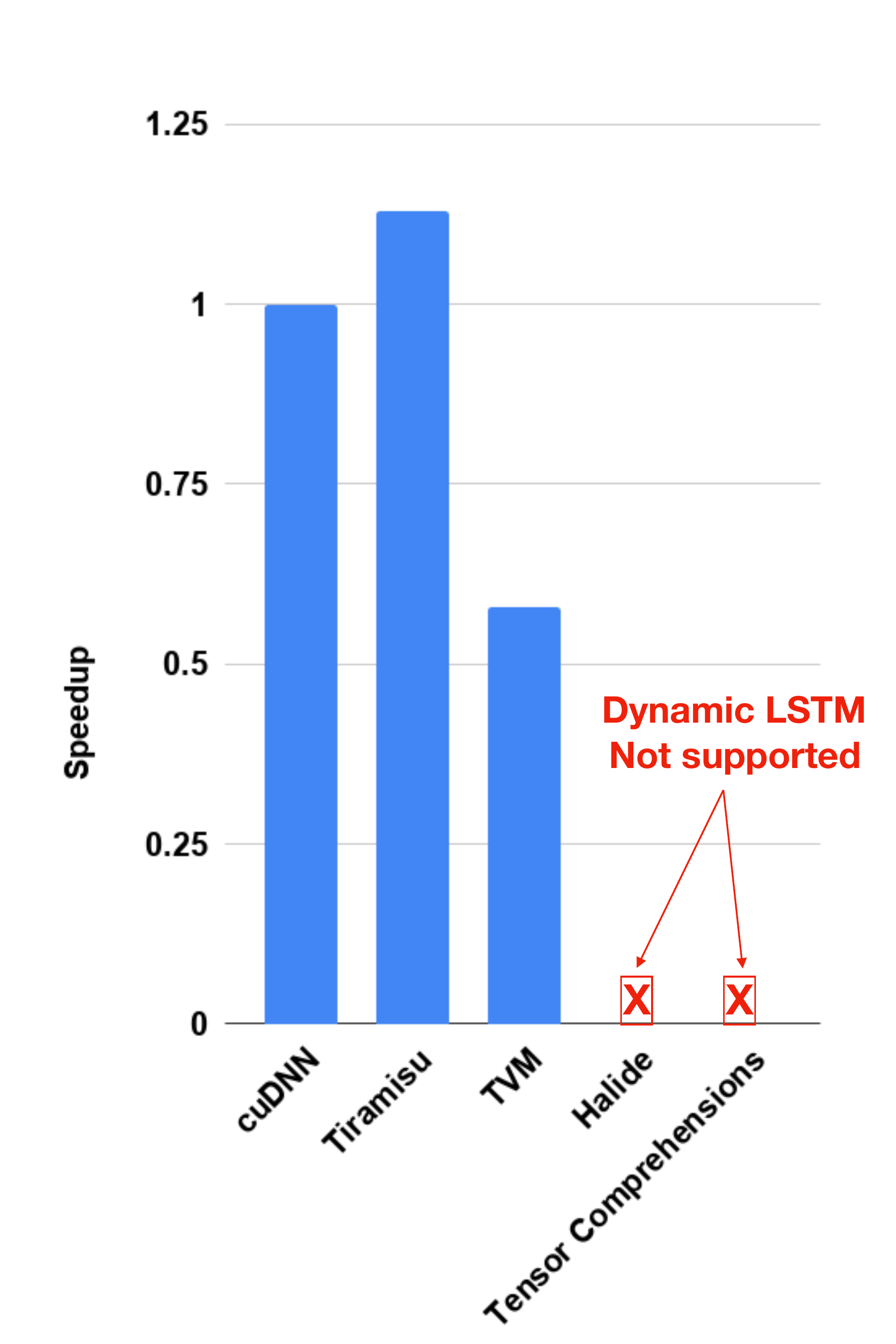}
        \vspace{-0.5cm}
        \caption{Speedups over cuDNN (dense).}
        \label{LSTMBenchmark}
    \end{minipage}
    \hspace{0.5cm}
    \begin{minipage}{0.14\textwidth}
        \vspace{1cm}
        \includegraphics[scale=0.17]{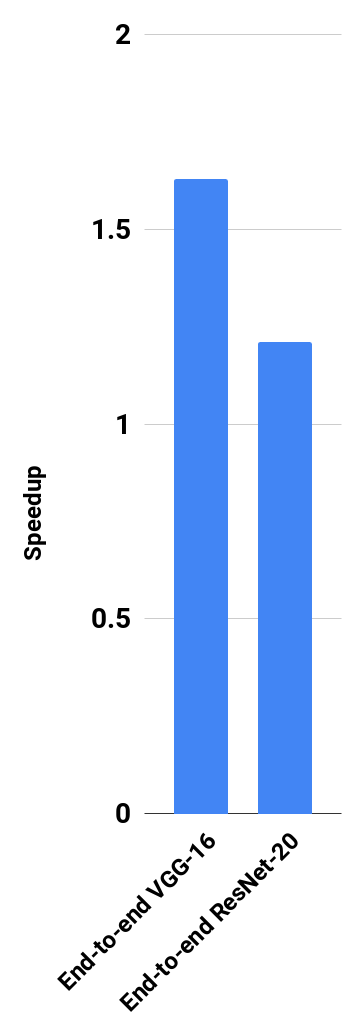}
        \vspace{-0.5cm}
        \caption{Speedups over MKL-DNN.}
        \label{endtoendBenchmark}
    \end{minipage}
  \vspace{-0.5cm}
\end{figure}

\begin{figure}[h!]
\centering
  \begin{minipage}{0.5\textwidth}
        \includegraphics[width=\columnwidth]{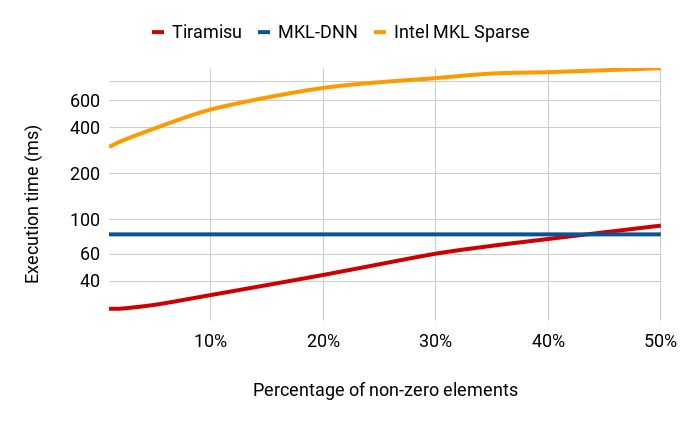}
        \caption{Break-even point for sparse convolution.\label{breakeven}}
  \end{minipage}
  \begin{minipage}{0.49\textwidth}
    \tiny
    \centering
    \setlength\tabcolsep{3pt}
    \begin{tabular}{l|l|l|l|l|}
        \hline
        
        \textbf{Feature} & \textbf{Tiramisu} & \textbf{TC} & \textbf{TVM} & \textbf{Halide} \\\hline

        \textbf{Express dynamic RNNs} & \yes & \no & \yes & \no  \\\hline

        \textbf{Optimize dynamic RNNs} & \yes & \no & \no & \no  \\\hline

        \textbf{Express/optimize sparse DNNs} & \yes & \no & \no & \no \\\hline

        \textbf{Generate distributed Code} & \yes & \no & \no & \yes \\\hline

        \textbf{Scheduling language} & \yes & \no & \yes & \yes  \\\hline

        \textbf{Support all affine transformations} & \yes & \yes & \no & \no  \\\hline





    \end{tabular}
    \caption{Comparison of DNN compilers\label{tab:related}}
  \end{minipage}
\end{figure}

The deep learning benchmarks include \texttt{Conv} (a direct implementation of a neural network convolution layer), \texttt{Conv-Relu-Maxpool} (a block of three layers, a direct convolution followed by a rectified linear unit followed by maxpooling), \texttt{Resize-Conv-Relu-Maxpool} (the same benchmark as the previous one but preceded by an image resizing step for preprocessing), \texttt{VGG} (a block of the VGG neural network~\cite{simonyan2014very}), \texttt{ResNet} (a block of the ResNet neural network~\cite{DBLP:journals/corr/HeZRS15}), and \texttt{Seq-to-seq} (a multilayer-LSTM that translates a sequence to another sequence~\cite{DBLP:journals/corr/SutskeverVL14}).

The use of a sparse convolution is not always profitable. Above certain density levels, a dense convolution implementation is more profitable than the sparse counterpart due to the overhead that the sparse implementation adds. Figure~\ref{breakeven} shows the break-even density level ($43.5\%$) after which a dense convolution implementation is faster than its sparse counterpart. The Intel MKL sparse implementation relies on sparse matrix multiplication and is slower than both implementations mainly due to the extra cost of lowering~\cite{park2016faster}.

The \texttt{VGG-Block} and \texttt{ResNet-Block} benchmarks in Figure~\ref{DNNBenchmark} are two representative blocks from the \texttt{VGG}~\cite{simonyan2014very} and \texttt{ResNet}~\cite{DBLP:journals/corr/HeZRS15} neural network architectures (a block is a repetitive sequence of layers in the neural network). We use the same sizes and parameters as in the original architectures. The sparse weights are obtained by applying the LTH pruning technique~\cite{frankle_stabilizing_2019}.
The blocks are chosen to be representative: first we exclude all the blocks that have a density level above $43.5\%$ and which should have a dense implementation; then, we compute the median of the weight density of the the remaining blocks; the chosen blocks have a density that is the closest to the median density.
Based on this methodology, we find that block 10 in both \texttt{ResNet} and \texttt{VGG} has the median density level (as shown in Table~\ref{tab:layerwise-densities}). The density level for block 10 is $16.1\%$ in \texttt{ResNet} and $1\%$ in \texttt{VGG}.
For \texttt{seq-to-seq}, we use the same architecture and sizes used in~\cite{216077} (4 LSTM layers, 100 elements in the input sequence and 1024 hidden parameters), and use $15\%$ as a uniformly distributed density level~\cite{kalchbrenner2018efficient}.

Figure~\ref{DNNBenchmark} shows a comparison between the performance of code generated by \framework{} (multicore CPU) and reference DNN libraries and frameworks. The baseline is the Intel MKL-DNN library (dense). The comparison includes the \framework{} implementation for dense weights, the \framework{} implementation for sparse weights, an implementation using Intel MKL sparse and the TensorFlow framework.

\framework{} outperforms the highly optimized Intel MKL-DNN library by up to $3\times$ in \texttt{Conv-Relu-Maxpool} and \texttt{Resize-Conv-Relu-Maxpool} due to operator fusion. \framework{} fuses the operators \texttt{Conv}, \texttt{Relu}, {Maxpool} (and \texttt{resize}) whereas Intel MKL-DNN has an implementation where only \texttt{Conv} and \texttt{Relu} are fused. For the sparse implementation, \framework{} outperforms the Intel MKL-DNN implementation by up to $5\times$. In \texttt{Conv-Relu-Maxpool} and \texttt{Resize-Conv-Relu-Maxpool}, in addition to the sparse implementation, we apply operator fusion. Figure~\ref{endtoendBenchmark} shows end-to-end speedups for sparse \framework{} compared to MKL-DNN (dense).

\vspace{-0.25cm}
\paragraph{LSTM Optimization on GPU}
Figure~\ref{LSTMBenchmark} compares the \framework{} GPU implementation of the \texttt{seq-to-seq} neural network with that of the cuDNN library~\cite{cudnn}, TVM, Halide and Tensor Comprehensions. While \framework{} and cuDNN use iteration space skewing to parallelize the multi-layer LSTM and increase the occupancy of the GPU, the TVM implementation does not support iteration space skewing and thus suffers from lower GPU occupancy. Halide and Tensor Comprehensions do not support dynamic LSTMs. In addition to the use of iteration space skewing to parallelize the \texttt{seq-to-seq} benchmark, the \framework{} implementation fuses multiple matrix multiplications into fewer multiplications to increase the GPU occupancy and uses the CUDA streams API to achieve concurrency on multiple GPUs~\cite{akkas2019}. \framework{} is faster than cuDNN in particular, because \framework{} tunes the number of fused matrix multiplications while knowing the size of the matrix multiplication whereas cuDNN does not provide such capability. In a separate experiment, we have found that the optimal number of fused matrix multiplication depends on the size of the LSTM matrix multiplication operations therefore.

\vspace{-0.25cm}
\section{Related Work\label{related}}
\vspace{-0.25cm}

Tensor Comprehensions~\cite{Vasilache2018TensorCF} and Diesel~\cite{Elango:2018:DDL:3211346.3211354} are fully automatic polyhedral compilers for deep learning designed mainly to target GPUs.
Unlike Tensor Comprehensions and Diesel, \framework{} has a scheduling language and therefore allows the user to have fine grain control over optimizations.
TVM~\cite{zhang2019deep} is another DNN compiler designed for targeting multiple hardware architecture. It has a scheduling language and uses machine-learning-based auto-tuning. TVM is not polyhedral though. It uses intervals to represent loop bounds and loop transformations which prevents TVM from applying certain transformations such as iteration space skewing (which is necessary for optimizing RNNs such as multilayer-LSTMs and increase GPU occupancy).
Other  machine  learning  domain  specific  compilers include TensorFlow XLA~\cite{45381}, DLVM~\cite{DBLP:journals/corr/abs-1711-03016}, Latte~\cite{Truong:2016:LLC:2980983.2908105} and SWIRL~\cite{doi:10.1177/1094342019866247}. Among all of the previous compilers, \framework{} is the only compiler that supports sparse DNNs.
Figure~\ref{tab:related} shows a comparison with some of these compilers
(\texttt{TC} in the table stands for Tensor Comprehensions).

Polyhedral compilers such as PENCIL~\cite{pencil,pencil_paper}, Pluto~\cite{bondhugula_practical_2008}, Polly~\cite{polly}, and PolyMage~\cite{Mullapudi:2015:PAO:2786763.2694364} are fully automatic.
While such fully automatic compilers provide productivity, they may not always obtain the best performance.  This is due to many reasons: these compilers do not implement some key optimizations such as array packing~\cite{Goto:2008:AHM:1356052.1356053}, register blocking, data prefetching (which are all supported by \framework{}). Besides, they do not have a precise cost-model to decide which optimizations are profitable.  For example, the Pluto~\cite{bondhugula_practical_2008} automatic scheduling algorithm (which is used for automatic scheduling in Pluto, PENCIL, Polly, and Tensor Comprehensions) tries to minimize the distance between producer and consumer statements while maximizing outermost parallelism, but it does not consider the data layout, redundant computations, or the complexity of the control of the generated code.  
Instead of fully automatic scheduling, \framework uses a more pragmatic approach and relies on a set of scheduling commands, giving the user full control over scheduling.

Other polyhedral compilers such as AlphaZ~\cite{yuki2012alphaz}, CHiLL~\cite{chill,Hall2010}, URUK~\cite{Girbal2006}, and Transformation Recipes~\cite{Hall:2009:LTR:2155247.2155251} allow users to express high-level transformations using scheduling commands. 
Since these frameworks are polyhedral, they can express any affine transformation. Their scheduling languages though only implement a subset of the transformations that are necessary to get peak performance. For example, they do not implement optimizations such as array packing, prefetching and register blocking.

Halide~\cite{halide_12} is an image processing DSL that has a scheduling language; however, it uses intervals to represent iteration spaces instead of the polyhedral model.  This limits the expressiveness of Halide.
For example, unlike \framework{}, Halide cannot naturally represent non-rectangular iteration spaces. It also cannot perform many complex affine transformations, such as iteration space skewing which is necessary for optimizing RNNs. In addition, Halide assumes that the program has an acyclic dataflow graph in order to simplify checking the legality of a schedule. This prevents users from expressing many programs with cyclic dataflow; for example, Halide does not allow the fusion of two loops (using the \texttt{compute\_with} command) if the second loop reads a value produced by the first loop. While this rule avoids illegal fusion, it prevents fusing many legal common cases. \framework{} avoids over-conservative constraints by relying on dependence analysis to check for the correctness of code transformations, enabling more possible schedules.


Exploiting sparsity in deep neural networks has been the subject of multiple projects. Park et al.~\cite{park2016faster} presented a fast algorithm for implementing sparse direct convolutions (on which we based our implementation), whereas Xuhao Chen~\cite{DBLP:journals/corr/abs-1802-10280}
Parashar et al.~\cite{DBLP:journals/corr/abs-1708-04485} on the other hand presented a hardware accelerator for sparse CNNs.

Acorns~\cite{acorns} is a framework designed mainly to optimize DNNs with input sparsity. It has a set of template codes for neural network operators and does not implement advanced loop nest optimizations such as iteration space skewing. Acorns introduces a data layout that exploits the structure of sparsity of input data in certain domains (LiDAR, face detection, character recognition, ...) where only certain specific regions of the input are non-zero. Unlike Acorns, \framework{} focuses on sparsity in weights.

\vspace{-0.25cm}
\section{Conclusion}
\vspace{-0.25cm}

In this paper, we demonstrate a DNN compiler that has two unique features: (1) it can generate efficient code for sparse DNNs; (2) it can optimize dynamic RNNs.
\framework can apply complex loop transformations thanks to the use of the polyhedral representation; and it relies on the use of scheduling commands, therefore it allows fine control over which optimizations to apply which allows \framework to reach high performance.
We evaluate \framework by implementing a set of deep learning benchmarks and show that \framework matches and outperforms the Intel MKL-DNN and cuDNN libraries by up to $5\times$ and outperforms state-of-the-art compilers by up to $2\times$.

\newpage

\bibliography{bibliography}

\end{document}